\documentclass[12pt]{article}

\begin{document}

\begin{titlepage}
\begin{flushright}
{\large \bf UCL-IPT-97-13}
\end{flushright}
\vskip 2cm
\begin{center}
\vskip .2in

{\Large \bf
On the Chern-Simons Topological Term at Finite Temperature}
\vskip .4in

{\large R. Gonz\'alez Felipe}\\[.15in]

{\em Institut de Physique Th\'eorique}\\

{\em Universit\'e catholique de Louvain}\\

{\em     
B-1348 Louvain-la-Neuve, Belgium}

\end{center}

\vskip 2in

\begin{abstract}
The  parity-violating topological term in the effective action  for 2+1 massive 
fermions is computed at finite temperature in the presence of a constant background  
field strength tensor. Gauge invariance of the finite-temperature effective action is
also discussed.
\end{abstract}
\end{titlepage}

A remarkable property of three dimensional gauge theories coupled to matter
is the dynamical generation of the Chern-Simons (CS)\ topological term\cite
{jackiw} through fluctuations of fermionic fields\cite{niemi,redlich,witten}%
. Non-Abelian CS theories are relevant for their intrinsic topological
structure, whereas the interest in the Abelian case stems from the fact that
they lead to fractional statistics in 2+1 dimensions. Thus, it is not
surprising that such an interaction turns out to be important both in the
context of topological field theories and in condensed matter physics, e.g.
in phenomena such as the fractional quantum Hall effect and high $T_{c}$
superconductivity.

It is well known that the classical non-Abelian CS action $S_{CS}$ is not
invariant under a homotopically nontrivial gauge transformation which
carries non-vanishing winding number. Requiring $\exp (iS_{CS})$ to be gauge
invariant leads to the discretization of the CS coefficient\cite{deser} at
the classical and quantum levels. On the other hand, for an Abelian CS\
theory, the CS coefficient remains arbitrary due to the absence of a similar
topological structure. In what concerns the effective gauge field action $%
S_{eff}[A]$ (obtained by integrating out the fermionic degrees of freedom),
it contains a gauge non-invariant piece which is however exactly cancelled
by the parity-violating term arising in any gauge invariant regularization
of the fermionic determinant\cite{redlich}. 

At finite temperature (or equivalently, when the time component is
compactified into a circle) the situation seems to be less understood.
Perturbative calculations reveal effective Chern-Simons actions with CS
coefficients which are smooth functions of the temperature\cite{niemi85}-%
\cite{aitchinson}. On the contrary, in ref.\cite{pisarski} it was argued
that the CS coefficient should be independent of the temperature. This
question has been recently addressed in refs.\cite{bralic,cabra}, where the
authors concluded, on gauge invariance grounds, that the coefficient of the
CS term at finite temperature cannot be smoothly renormalized and that it is
at most an integer-valued function of the temperature. Nevertheless,
calculations of the latter coefficient for an effective action of a massive
fermion system in 0+1 dimensions\cite{dunne} and in the 2+1 Abelian and
non-Abelian cases\cite{griguolo,fosco}, have shown that the complete effective action
can be made gauge invariant. Therefore, although the perturbative result
yielding a smooth dependence of the CS coefficient on the temperature is
correct, any perturbative order is not sufficient to preserve the gauge
invariance.

In this letter we shall follow the approach originally used in 2+1 \textit{%
QED}\ at zero temperature\cite{redlich} and compute the parity-violating
contribution to the fermionic current at finite temperature for the special
case of gauge fields which produce a constant field strength tensor $F_{\mu
\nu }.$ The present calculation confirms the fact that gauge invariance of
the finite-$T$ Chern-Simons effective action holds for an arbitrary winding
number, provided the induced (mass and temperature-dependent) parity
breaking contribution as well as the (temperature-independent) parity
anomalous contribution are taken into account\cite{griguolo,fosco}. In the first
order of perturbation theory, the usual perturbative expression for the CS
term at finite temperature is also reproduced.

We start by defining the three-dimensional (Euclidean) effective action $%
S_{eff}[A,m]$ for a massive fermionic field $\psi (\tau ,x)$ in a gauge
background field $A_{\mu }(\tau ,x),$ with the time direction compactified
into the interval $0\leq \tau \leq \beta =1/T,$

\begin{equation}
e^{-S_{eff}[A,m]}=\int \mathcal{D}\psi \mathcal{D}\bar{\psi}\exp \left(
-\int_{0}^{\beta }d\tau \int d^{2}x\; \bar{\psi}(\not{\!\partial}+ie\not%
{\!\!A}+m)\psi \right) \ ,  \label{1}
\end{equation}
where $T$ is the temperature and the Euclidean Dirac matrices are taken in
the representation $\gamma _{0}=\sigma _{3},\gamma _{1}=\sigma _{1},\gamma
_{2}=\sigma _{2};$ $\sigma _{\mu }$ are the Pauli matrices. The functional
integral in Eq.(\ref{1}) must be evaluated using periodic (antiperiodic) in $%
\tau $ boundary conditions for the gauge (fermionic) fields, i.e.
\begin{eqnarray}
A_{\mu }(\beta ,\mathbf{x}) &=&A_{\mu }(0,\mathbf{x})\ ,  \label{2} \\
\psi (\beta ,\mathbf{x}) &=&-\psi (0,\mathbf{x})\ .  \label{3}
\end{eqnarray}

The allowed non-trivial gauge transformations at finite temperature are
those which preserve the above conditions, namely,

\begin{eqnarray}
\psi (\tau ,\mathbf{x}) &\rightarrow &e^{-ie\vartheta (\tau ,\mathbf{x}%
)}\psi (\tau ,\mathbf{x})\ ,  \nonumber \\
\bar{\psi}(\tau ,\mathbf{x}) &\rightarrow &e^{ie\vartheta (\tau ,\mathbf{x})}%
\bar{\psi}(\tau ,\mathbf{x})\ ,  \label{4} \\
A_{\mu }(\tau ,\mathbf{x}) &\rightarrow &A_{\mu }(\tau ,\mathbf{x})+\partial
_{\mu }\vartheta (\tau ,\mathbf{x})\ ,  \nonumber
\end{eqnarray}
with
\begin{equation}
\vartheta (\beta ,\mathbf{x})=\vartheta (0,\mathbf{x})+\frac{2\pi }{e}n\ ,
\label{5}
\end{equation}
where the integer number $n$ characterizes the homotopy class of the gauge
transformation.

In what follows we restrict ourselves to gauge field configurations which
induce a constant field strength tensor $F_{\mu \nu }=\partial _{\mu }A_{\nu
}-\partial _{\nu }A_{\mu }$ and with trivial winding number $n=0$. We also
assume the component $A_{0}$ to be independent of $\mathbf{x}$. Then, the
remaining $\tau $ dependence in $A_{0}$ can always be removed by a
redefinition of the fermion fields and the proper choice of a function $%
\vartheta (\tau ).$ Indeed, from the equation $\partial _{\tau }\vartheta
(\tau )=-A_{0}(\tau )+\bar{A}_{0},$ with $\bar{A}_{0}$ a constant, and the
condition $\vartheta (\beta )=\vartheta (0),$ we obtain

\[
\vartheta (\tau )=-\int_{0}^{\tau }A_{0}(\tau )d\tau +\bar{A}_{0}\tau \
,\quad \bar{A}_{0}=\frac{1}{\beta }\int_{0}^{\beta }A_{0}(\tau )d\tau . 
\]

Under the above assumptions, the gauge field $A_{\mu }$ can be written as

\begin{equation}
A_{\mu }=(A_{0},\frac{1}{2}F_{ij}x_{j})\ ;\ A_{0}\equiv \bar{A}_{0};\quad
i,j=1,2\ .  \label{6}
\end{equation}
We note that this choice corresponds to a vanishing electric field ($%
F_{0i}=0)$ and a constant magnetic field. This gauge configuration will
allow us to compute exactly the Chern-Simons topological current at finite
temperature, following an approach similar to the zero-temperature case\cite
{redlich}.

The ground-state current in the presence of the background field (\ref{6})
is defined as
\begin{equation}
\left\langle J^{\mu }\right\rangle =\frac{\delta S_{eff}[A,m]}{\delta A_{\mu
}}\ ,  \label{7}
\end{equation}
where the total effective action is given, according to Eq.(\ref{1}), by

\begin{equation}
S_{eff}[A,m]=-\mbox{Tr}\ln (\not{\!\partial}+ie\not{\!\!A}+m)\ .  \label{8}
\end{equation}
Notice that since the only parity-odd term in the Euclidean effective action
is the mass term, we can obtain the CS\ current through the combination 
\begin{equation}
\left\langle J_{CS}^{\mu }\right\rangle =\frac{1}{2}\left( \frac{\delta
S_{eff}[A,m]}{\delta A_{\mu }}-\frac{\delta S_{eff}[A,-m]}{\delta A_{\mu }}%
\right) \ .  \label{9}
\end{equation}

The (gauge-invariant) background current (\ref{7}) can be written in terms
of the Green function for the Dirac operator\cite{schwinger},

\begin{equation}
\left\langle J^{\mu }\right\rangle =\left. e\mbox{Tr}\left[ \gamma ^{\mu
}\exp \left( -ie\int_{x}^{x^{\prime }}A(\xi )\cdot d\xi \right)
G(x,x^{\prime })\right] \right| _{x\rightarrow x^{\prime }}\ ,  \label{10}
\end{equation}
where 
\begin{equation}
G(x,x^{\prime })=\left\langle x\left| G\right| x^{\prime }\right\rangle
=\left\langle x\left| \left( \not{\!\Pi}+im\right) ^{-1}\right| x^{\prime
}\right\rangle \ ,  \label{11}
\end{equation}
$x^{\mu }\equiv (\tau ,\mathbf{x),}\Pi _{\mu }=p_{\mu }-eA_{\mu }$ and $%
\left[ \Pi _{\mu },\Pi _{\nu }\right] =ieF_{\mu \nu }.$ As customary at
finite temperature, calculations are carried out by restricting the values
of the Euclidean time $\tau $ to the interval $0\leq \tau \leq \beta =1/T.$
We have then 
\begin{equation}
G(x,x^{\prime })=\frac{1}{\beta }\sum_{n=-\infty }^{\infty }\int \frac{d^{2}p%
}{(2\pi )^{2}}G\ e^{ip(x-x^{\prime })},  \label{12}
\end{equation}
with $p_{0}=\omega _{n}=(2n+1)\pi /\beta $ - the Matsubara frequencies; $%
px=\omega _{n}\tau +\mathbf{px}$.

To calculate explicitly $\left\langle J_{CS}^{\mu }\right\rangle ,$ let us
write the operator $G$ in the equivalent form\cite{schwinger}

\begin{equation}
G=\frac{1}{\not{\!\Pi}+im}=\left( \not{\!\Pi}-im\right) \int_{0}^{\infty }ds\
e^{-s\left( \not{\Pi}^{2}+m^{2}\right) }.  \label{13}
\end{equation}
According to the definition given in Eq.(\ref{9}), it is easy to see that
the only contribution to the CS current comes from the term proportional to $%
im$ in Eq.(\ref{13}). Thus, we shall consider the simplified function

\begin{equation}
G_{CS}=\left. -\frac{im}{\beta }\sum_{n=-\infty }^{\infty }\int \frac{d^{2}p%
}{(2\pi )^{2}}\int_{0}^{\infty }ds\ e^{-s\left( \not{\Pi}^{2}+m^{2}\right)
}e^{ip(x-x^{\prime })}\right| _{x\rightarrow x^{\prime }}.  \label{14}
\end{equation}
Using the relation $\not{\!\!\Pi}^{2}=\Pi ^{2}+\frac{e}{2}\sigma _{\mu \nu
}F^{\mu \nu },\sigma _{\mu \nu }=\frac{i}{2}\left[ \gamma _{\mu },\gamma
_{\nu }\right] $, and the fact that $\left[ \Pi _{0},\Pi _{i}\right] =0$ for
the gauge configuration (\ref{6}), it is easy to factorize the $\tau $
dependence in Eq.(\ref{14}) to obtain

\begin{equation}
G_{CS}=\left. -\frac{im}{\beta }\sum_{n=-\infty }^{\infty }\int_{0}^{\infty
}ds\ e^{-s\left[ (\omega _{n}-eA_{0})^{2}+\frac{e}{2}\sigma \cdot
F+m^{2}\right] }\left\langle \mathbf{x}\left| e^{-s\mathbf{\Pi }^{2}}\right| 
\mathbf{x}^{\prime }\right\rangle \right| _{\mathbf{x}\rightarrow \mathbf{x}%
^{\prime }}.  \label{15}
\end{equation}

The matrix element appearing in the last equation can be evaluated using the
method developed by Schwinger\cite{schwinger} to perform similar
calculations in four-dimensional \textit{QED}. We define the evolution
operator $U(s)=e^{-s\mathcal{H}},\mathcal{H}=\mathbf{\Pi }^{2},$ which
describes the development in the proper (imaginary) time $s$ of a system
with Hamiltonian $\mathcal{H}$. Then

\begin{equation}
\left\langle \mathbf{x}\left| e^{-s\mathbf{\Pi }^{2}}\right| \mathbf{x}%
^{\prime }\right\rangle =\left\langle \mathbf{x}\left| U(s)\right| \mathbf{x}%
^{\prime }\right\rangle =\left\langle \mathbf{x},s\mid \mathbf{x}^{\prime
},0\right\rangle \ ,  \label{16}
\end{equation}
with the boundary condition $\left. \left\langle \mathbf{x,}s\mid \mathbf{x}%
^{\prime },0\right\rangle \right| _{s\rightarrow 0}=\delta (\mathbf{x-x}%
^{\prime }).$

Solving the associated dynamical problem

\begin{equation}
\frac{dx_{i}}{ds}=\left[ H,x_{i}\right] =-2i\Pi _{i}\mathbf{\ ,\ }\frac{d\Pi
_{i}}{ds}=\left[ H,\Pi _{i}\right] =-2ieF_{ij}\Pi _{j}\ ,  \label{17}
\end{equation}
finally we obtain

\begin{equation}
\left. \left\langle \mathbf{x},s\mid \mathbf{x}^{\prime },0\right\rangle
\right| _{\mathbf{x}\rightarrow \mathbf{x}^{\prime }}=\frac{1}{4\pi }\frac{e|%
\tilde{F}|}{\sinh (es|\tilde{F}|)}\ ,  \label{18}
\end{equation}
where $\tilde{F}^{\mu }=\frac{1}{2}\epsilon ^{\mu \alpha \beta }F_{\alpha
\beta }$ is the dual to $F_{\mu \nu }$ and $|\tilde{F}|=\sqrt{B^{2}-E^{2}}.$

In evaluating the Chern-Simons current (cf. Eq.(\ref{10})) we also need the
following trace formula:
\begin{equation}
\mbox{Tr}\left[ \gamma ^{\mu }e^{-\frac{es}{2}\sigma \cdot F}\right] =-2%
\frac{\tilde{F}^{\mu }}{|\tilde{F}|}\sinh (es|\tilde{F}|).  \label{19}
\end{equation}

From Eqs.(\ref{10}),(\ref{15}),(\ref{18}) and (\ref{19}) we find

\begin{eqnarray}
\left\langle J_{CS}^{\mu }\right\rangle &=&\frac{ime^{2}\tilde{F}^{\mu }}{%
2\pi \beta }\sum_{n=-\infty }^{\infty }\int_{0}^{\infty }ds\ e^{-s\left[
(\omega _{n}-eA_{0})^{2}+m^{2}\right] }  \nonumber \\
&=&\frac{ime^{2}\tilde{F}^{\mu }}{2\pi \beta }\sum_{n=-\infty }^{\infty }%
\frac{1}{(\omega _{n}-eA_{0})^{2}+m^{2}}\ .  \label{20}
\end{eqnarray}
The last sum in the above equation is easily evaluated and we arrive at our
final result:

\begin{equation}
\left\langle J_{CS}^{\mu }\right\rangle =\frac{ie^{2}}{4\pi }\frac{m}{|m|}%
\tilde{F}^{\mu }\frac{\sinh (\beta |m|)}{\cosh (\beta |m|)+\cos (e\beta
A_{0})}\ .  \label{21}
\end{equation}

First notice that in the limit $T\rightarrow 0$ we reproduce the standard
zero temperature result\cite{redlich}, namely,

\begin{equation}
\left\langle J_{CS}^{\mu }\right\rangle =\frac{ie^{2}}{4\pi }\frac{m}{|m|}%
\tilde{F}^{\mu }.  \label{22}
\end{equation}
We can now functionally integrate over the field $A_{\mu }$ in Eq.(\ref{20})
to obtain the Chern-Simons effective action

\begin{equation}
S_{CS}[A]=\frac{ie}{2\pi }\frac{m}{|m|}\arctan \left[ \tanh \left( \frac{%
\beta |m|}{2}\right) \tan \left( \frac{e\beta A_{0}}{2}\right) \right] \int
d^{2}x\epsilon _{ij}\partial _{i}A_{j}\ .  \label{23}
\end{equation}
This result coincides with the one recently obtained in refs.\cite{griguolo,fosco} by
calculating the fermion determinant. In particular, when $T\rightarrow 0$ it
gives the standard zero-temperature result 
\begin{equation}
S_{CS}[A]=\frac{ie^{2}}{8\pi }\frac{m}{|m|}\int d^{3}x\tilde{F}_{\mu }A_{\mu
}=\pm 2\pi W[A],  \label{24}
\end{equation}
where the $\pm $ sign depends on the sign of $m$ and $W[A]$ is the
Chern-Simons secondary characteristic class number\cite{deser},

\begin{equation}
W[A]=\frac{ie^{2}}{16\pi ^{2}}\int d^{3}x\tilde{F}_{\mu }A_{\mu }\ .
\label{25}
\end{equation}

We note that under non-trivial gauge transformations with winding number $%
n\neq 0,$ the argument of $\tan (e\beta A_{0}/2)$ in Eq.(\ref{23}) is
shifted by $n\pi $ and so does the value of the arctan function, once the
branch used in its definition is also properly shifted. Thus, a situation
similar to the $T=0$ case is occurring, namely, the gauge non-invariance of
the Chern-Simons effective action is compensated by the parity anomalous
contribution\cite{redlich}, which arises in any gauge-invariant
regularization scheme and amounts to $\pm 2\pi W[A]$.

Eq.(\ref{23}) also reproduces in the first order of perturbation (in $e$)
theory, the usual perturbative expression:

\begin{equation}
S_{CS}[A,m]=\pm 2\pi \tanh \left( \frac{\beta |m|}{2}\right) W[A]\ .
\label{26}
\end{equation}
This perturbative result shows however an apparent breaking of gauge
invariance: the gauge non-invariance present in the effective action (\ref
{26}) is no longer cancelled by the (temperature-independent) parity
anomaly. From this we can conclude that any perturbative order is not
sufficient to preserve gauge invariance and thus, a full (non-perturbative)
answer is necessary in order to discuss gauge invariance at finite
temperature. 

Finally, let us briefly comment on the non-Abelian case. Similar
calculations can be performed for $SU(N)$ non-Abelian gauge-field
configurations which induce a constant field strength tensor $F_{\mu \nu
}^{a}$ ($a$ is the colour index). As is well-known, there are only two types
of such fields\cite{redlich}. The first one is an Abelian-type field, which
following arguments similar to the ones presented here(see Eq.(\ref{6})) can
be written in the form $A_{\mu }^{a}=(\eta ^{a}A_{0},\frac{1}{2}\eta
^{a}F_{ij}x_{j}),$ where $\eta ^{a}$ is a constant unit vector in colour
space. The second one is a genuine non-Abelian constant gauge field, $A_{\mu
}^{a}=const$, with $[A_{\mu },A_{\nu }]\neq 0,A_{\mu }\equiv A_{\mu
}^{a}T_{a}$ and $T_{a}$ the generators of the Lie algebra. For such fields,
the gauge invariance of the effective Chern-Simons action under non-trivial
gauge transformations is a more subtle\cite{fosco} question which deserves
further study.

\noindent {\large Acknowledgements}

I am grateful to D. Del\'{e}pine, P. Ruelle and J. Weyers for valuable
discussions and comments.

\end{document}